\documentclass[aps,prl,twocolumn,groupedaddress]{revtex4}

\usepackage[dvips]{graphics, color}

\input epsf

\begin{document}

\newcommand{\ra}{$^{85}$Rb }
\newcommand{\rb}{$^{87}$Rb }
\newcommand{\fr}{Feshbach resonance }

\title{Observation of Heteronuclear Feshbach Molecules from a $^{85}$Rb -- $^{87}$Rb gas}
\author{S. B. Papp}
\email[Email: ]{papp@jilau1.colorado.edu}
\author{C. E. Wieman}
\address{JILA, National Institute of Standards and Technology
and University of Colorado, Boulder, Colorado 80309-0440, USA}
\date{\today}

\begin{abstract}
We report on the observation of ultracold heteronuclear Feshbach molecules.  Starting with a \rb
BEC and a cold atomic gas of $^{85}$Rb, we utilize previously unobserved interspecies Feshbach
resonances to create up to 25,000 molecules.  Even though the \ra gas is non degenerate, we observe
a large molecular conversion efficiency due to the presence of a quantum degenerate \rb gas; this
represents a key feature of our system.  We compare the molecule creation at two different Feshbach
resonances with different magnetic-field widths.  The two Feshbach resonances are located at
$265.44\pm0.15$ G and $372.4\pm1.3$ G. We also directly measure the small binding energy of the
molecules through resonant magnetic-field association.
\end{abstract}


\maketitle

The creation of ultracold molecules from ultracold atoms is currently a topic of great experimental
and theoretical interest \cite{Donley2002a,Regal2003c,Sage2005a,Kohler2006a}. Ultracold
heteronuclear molecules in low-lying vibrational states are particularly interesting since they are
predicted to exhibit a permanent dipole moment due to the unequal distribution of electrons.
Numerous proposals for utilizing polar molecules exist, including quantum computation
\cite{Demille2002a} and the search for the electron electric dipole moment \cite{Hudson2002a}.
Although no significant permanent dipole moment is expected to exist in a $^{85}$Rb--$^{87}$Rb
molecule, our work demonstrates a first step toward the efficient production of ground-state
ultracold heteronuclear molecules.

To date cold heteronuclear molecules in high-lying vibrational levels have been created using
photoassociation \cite{Kerman2004a,Wang2004a,Mancini2004a}. These molecules can then be pumped
toward low-lying vibrational levels by exciting bound-bound molecular transitions \cite{Sage2005a}
via, for example, stimulated Raman-type transitions \cite{Demille2002a} that enhance the
probability of populating the lowest vibrational level. The initial photoassociation step used in
this process is inefficient, and many final vibrational levels of the molecule are occupied. An
alternative to this initial photoassociation step is the direct conversion of two free atoms into a
molecule in the highest vibrational levels using a Feshbach resonance
\cite{Donley2002a,Regal2003c,Durr2003a,Herbig2003a,Cubizolles2003a,Jochim2003a,Strecker2003a,Thompson2005b}.
High conversion efficiency using a Feshbach resonance has been demonstrated in single-species gases
via adiabatic magnetic-field sweeps across the resonance \cite{Hodby2005a,Kohler2006a}, three-body
recombination \cite{Jochim2003a}, resonant magnetic-field association \cite{Thompson2005b}, and non
adiabatic magnetic-field sweeps \cite{Donley2002a,Claussen2002a}.  Feshbach resonances between two
different atomic species \cite{Inouye2004a,Stan2004a} have previously been reported. Our work
builds upon these observations by demonstrating stable heteronuclear Feshbach molecules.

In this Letter we present a systematic study of the creation of heteronuclear molecules from an
atomic gas of \ra and $^{87}$Rb. We find that these molecules can be created using the standard
techniques already proven with single-species gases.  The presence of two species with different
quantum degeneracy provides a rich system for testing our understanding of the conversion
efficiency from atoms to molecules.  Furthermore, the molecule creation process allows us to
determine the location and width of Feshbach resonances in the two-species system; this information
will be required for future studies of the $^{85}$Rb--$^{87}$Rb system with a tunable interspecies
interaction.

Our apparatus is designed to create a two-species Bose gas through sympathetic cooling of \ra with
$^{87}$Rb.  The details of our system are similar to those described in Refs.
\cite{Cornish2000a,Lewandowski2003a} and are only briefly described here. We initially collect
approximately $3\times10^9$ $^{87}$Rb atoms and $10^7$ $^{85}$Rb atoms in a two-species vapor cell
magneto-optical trap (MOT). The atoms are loaded into a quadrupole magnetic trap with a 100 G/cm
magnetic-field gradient in the axial direction of the coils and physically transported to another
region of the apparatus with lower vacuum pressure \cite{Greiner2001a,Lewandowski2003a}. Here the
atoms are prepared in the $^{85}$Rb $|f=2, m_f=-2\rangle$ state and $^{87}$Rb $|f=1, m_f=-1\rangle$
state and loaded into a Ioffe-Pritchard-type magnetic trap. Selective rf evaporation is used to
lower the temperature of the $^{87}$Rb gas, while the $^{85}$Rb gas is sympathetically cooled
through thermal contact with the $^{87}$Rb \cite{Bloch2001a}.

\begin{figure}[htb]
\begin{center}
\scalebox{0.35}[0.35]{\includegraphics{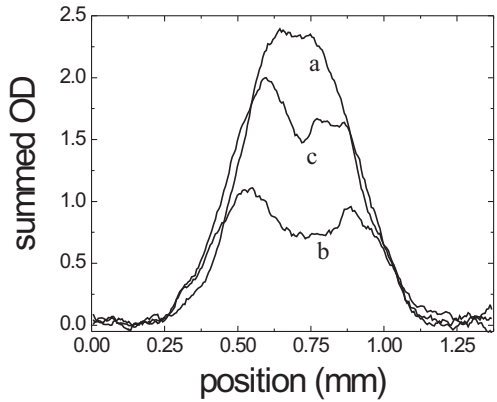}} \caption{
Absorption-image axial cross sections of the \ra gas demonstrating
reversible molecule creation. The measured two-dimensional optical
density (OD) was summed in the remaining radial direction of the
absorption image. (a) Prior to sweeping the magnetic field the
atom number is 26,500. (b) After sweeping through the resonance
53\% of the gas is converted to molecules. (c)  By reversing the
molecule creation process approximately 85\% of the initial atom
number is observed to reappear. Note that most of the \ra loss
occurs in the center of the gas where the \rb BEC density is
largest. \label{cloud}}
\end{center}
\end{figure}

The mixed-species gas is evaporatively cooled in the magnetic trap to approximately 10 $\mu$K and
then further evaporatively cooled in an optical dipole trap. The optical trap is formed at the
focus of a Yb:YAG laser beam with a $1/\, e^2$ radius of 23.5 $\mu$m. To evaporate in the optical
trap the power is lowered from 1.2 W to 14 mW in 10 seconds. This trap has the distinct advantage
of holding any spin state in the same spatial location while allowing a variable magnetic field to
be applied. A Helmholtz pair of coils provides a magnetic field up to 700 G; the magnetic field is
calibrated using rf-driven Zeeman transitions with a systematic uncertainty of 0.01$\%$. We perform
the optical trap evaporation slightly above the \ra Feshbach resonance at 155 G where
density-dependent $^{85}$Rb--$\,^{85}$Rb loss is minimized \cite{Roberts2000a}. We typically
produce a \rb BEC with 300,000 atoms and 50$\%$ condensate fraction and a non degenerate (T/T$_c =
2.4$) gas of \ra with 40,000 atoms. Measurements are performed in an optical trap with a radial
trap frequency of $\omega_r = 2\,\pi\times200$ Hz and an aspect ratio of approximately 100.

\begin{figure}[htb]
\begin{center}
\scalebox{0.8}[0.8]{\includegraphics{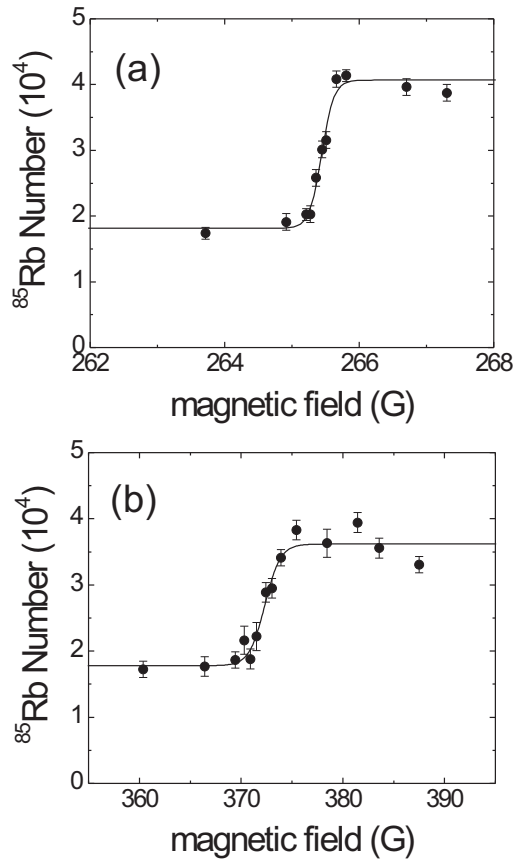}} \caption{Atom
loss after magnetic-field sweeps through a Feshbach resonance. The
number of \ra atoms is a function of the final magnetic field
during the sweeps near the (a) 265 G and (b) 372 G Feshbach
resonances.  Initially there are $2.1-2.3 \times 10^5$ \rb atoms
at T/T$_c$ between 0.82 and 0.84. The data are fitted to an error
function to extract the center position and width. The resulting
positions of the two transitions are $265.44\pm0.15$ G and
$372.4\pm1.3$ G with the uncertainty given by the fitted RMS
width. \label{pos}}
\end{center}
\end{figure}

We have discovered two heteronuclear $s$-wave Feshbach resonances in the $^{85}$Rb $|f=2,
m_f=-2\rangle$ state and $^{87}$Rb $|f=1, m_f=-1\rangle$ state, one near 265 G and the other near
372 G. The binding energy of the bound molecular state increases with magnetic field above the
position of each Feshbach resonance. To create heteronuclear molecules we adiabatically (450
$\mu$s/G) sweep \cite{Regal2003c} the magnetic field from low to high field through a Feshbach
resonance. Figure \ref{cloud}b shows an absorption-image cross section of the \ra gas following a
sweep through the resonance.  Prior to imaging the gas is released from the optical trap and
expands for 3 ms; at the same time as the optical trap release is initiated, the magnetic field is
switched off within 50 $\mu$s. A significant fraction ($\sim60\%$) of the atoms disappear as
compared to an absorption-image cross section without sweeping the magnetic field (Fig.
\ref{cloud}a). When we reverse the molecule creation process by immediately applying a second field
sweep in the opposite direction less than 100 $\mu$s after the first sweep ends, (Fig.
\ref{cloud}c), a large fraction of the atoms reappear \cite{Regal2003c}.  The atoms that reappear
represent reversible heteronuclear molecule formation. The small fraction of atoms that do not
return are lost to $^{85}$Rb--$\,^{87}$Rb inelastic collisions that remove atoms from the trap.  We
verified this by sweeping across the resonance at the same rate in the opposite direction. At a
typical \rb peak density of $1\times10^{14}$ cm$^{-3}$ we observe approximately $15\%$ \ra number
loss during a sweep. By monitoring the reappearance of atoms as a function of time, we find that
the molecules decay with a lifetime of approximately 1 ms.  This decay is due to a combination of
inelastic collisions with the \rb gas and a one-body spontaneous process
\cite{Thompson2005a,Kohler2005a}.

We precisely located the Feshbach resonances with an experimental technique that avoids the need
for rapid magnetic-field sweeps. At the lower (higher) field resonance we sweep the magnetic field
from 269 G (397 G) downward toward the resonance stopping at various final values.  The rate of
this magnetic field sweep is fast compared to both the timescale for molecule creation and atom
loss due to inelastic collisions. The field is held at the final value for 0.3 ms and then returns
to 269 G (397 G) at a rate of 450 $\mu$s/G (70 $\mu$s/G) which is slow compared to the molecule
creation rate. The magnetic field remains here for 5 ms to ensure that any molecules made during
the second sweep decay and are lost from the optical trap \cite{Thompson2005a,Kohler2005a}. We then
simultaneously turn off the magnetic field and the optical trap to let the gas expand for 6 ms and
measure the number of \ra atoms remaining. In Fig. \ref{pos} we show the atom number remaining as a
function of the final magnetic field for the two Feshbach resonances.  The key to this method is
that if the field passes through the Feshbach resonance on the first sweep, then a fraction of the
atoms will be converted into molecules by the second sweep. The rapid onset of atom loss due to
molecule creation when the magnetic field is swept below 265.44 G (372.4 G) represents crossing the
peak of the Feshbach resonance. A significant fraction ($\sim50\%)$ of the \ra gas is converted to
molecules.

The dependence of molecule conversion efficiency on sweep rate was measured in Ref.
\cite{Hodby2005a} and is well characterized by a Landau-Zener model. For our two-species gas with
similar number and temperature we observe a factor of $5.3\pm1.9$ decrease in the ramp rate
required to create molecules at the 372 G Feshbach resonance as compared to the 265 G resonance.
This factor is consistent with the ratio of the predicted widths of the two resonances, which is in
the range of 6--8 \cite{Burke1998a,Kokkelmanns2006a,Bohn2006a}. This directly verifies the
predicted inverse relationship between the Feshbach resonance width and the sweep rate required to
create molecules.

    \begin{figure}[htb]
    \begin{center}
    \scalebox{0.5}[0.5]{\includegraphics{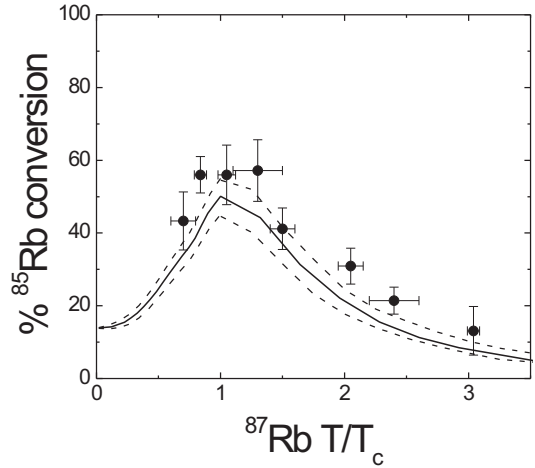}}
    \caption{
Heteronuclear molecule conversion efficiency at the 265 G Feshbach
resonance as a function of \rb T/T$_c$. At our largest conversion
efficiency the \ra gas has T/T$_c=2.6$. The solid line shows a
simulation based on our conversion model, and the dashed lines
represent the uncertainty. The conversion drops below T/T$_c=1$
since the BEC is spatially smaller than the thermal \ra gas and
the conversion process depends on the proximity of two atoms in
phase space.
    \label{conv}}
    \end{center}
    \end{figure}

We have investigated the adiabatic conversion efficiency of atoms to molecules. We begin by
sweeping the magnetic field upward through the 265 G Feshbach resonance at a rate that is slow with
respect to the molecule conversion rate.  It is then held at 269 G for 5 ms to allow the molecules
to decay and be lost from the trap. The number of molecules formed is simply the difference in \ra
number before and after the sweep with a small correction applied to account for measured inelastic
atom loss during the field sweeps. The conversion efficiency to heteronuclear molecules is shown in
Fig. \ref{conv} as a function of \rb T/T$_c$.  We varied T/T$_c$ by either changing the number of
\rb atoms or the temperature. The \ra gas had a T/T$_c$ in the range of 2.2 to 4. We observe up to
$60\%$ conversion of the \ra gas into molecules even when the \ra gas is far from quantum
degeneracy. The largest conversion efficiency is observed when \ra is least quantum degenerate,
indicating that the conversion efficiency primarily depends on \rb T/T$_c$.

In a single-species gas the molecule conversion efficiency of an adiabatic field sweep was shown to
depend only on the degree of quantum degeneracy \cite{Hodby2005a,Williams2005a}.  A molecule is
formed during an adiabatic sweep if two free atoms are sufficiently close in phase space so that
their wavefunction can smoothly evolve to a bound molecule as the Feshbach resonance is crossed.
This model also explains the behavior we observe here. Each particle of the less degenerate species
is surrounded in phase space by particles of the more degenerate \rb gas with which molecules can
be formed. We used a Monte-Carlo simulation to model the observed conversion efficiency. Phase
space distributions of position and momenta of \ra and \rb are randomly generated based on the
number and temperature of each gas and the trap frequencies. For the \rb gas at temperatures above
T$_c$ a Maxwell-Boltzmann (MB) distribution is used. Below T$_c$ we use a combination of a
Thomas-Fermi distribution for the BEC and a Bose-Einstein distribution for the thermal component.
A MB distribution is always used for the \ra gas since the temperature of the gas does not fall
below T $= 2.2 $\,T$_c$. For each \ra atom the simulation searches the \rb gas to find a pair that
is sufficiently close in phase space. After a pairing occurs, the two atoms are removed from the
simulation. Two atoms are considered sufficiently close in phase space if the following conditions
are met: (1) If a BEC is present, any \ra atoms inside the Thomas Fermi radius of the condensate
form a molecule, and (2) outside the BEC, an \ra atom and a partner \rb atom must satisfy the
relation found in Ref. \cite{Hodby2005a}: $|\delta r_{rel}\, m\, \delta v_{rel}|<\gamma\,h$, where
$\delta r_{rel}$ is the separation of the pair, $m$ is the atomic mass, $\delta v_{rel}$ is the
relative velocity, and $\gamma = 0.44\pm0.03$.

The results of our simulation are shown by the solid line in Fig. \ref{conv}. For T/T$_c>1$ each
gas has roughly the same spatial size in the optical trap, and therefore molecule conversion can
occur anywhere in the gas as long as the local phase space criterion is met.  At T/T$_c=1$ and
below, a significant fraction of the \rb atoms are part of the condensate, which is spatially small
compared to the extent of the \ra gas.  If the \ra gas were a BEC, our simulation would predict
that the molecule conversion efficiency quickly approaches 100\% as the T/T$_c$ of each gas drops
below one.  There is good agreement between the experiment and simulation in both the non
degenerate and quantum degenerate regimes, suggesting that the pairing model in Ref.
\cite{Hodby2005a} is also applicable to heteronuclear molecule creation.

\begin{figure}[htb]
    \begin{center}
    \scalebox{0.4}[0.4]{\includegraphics{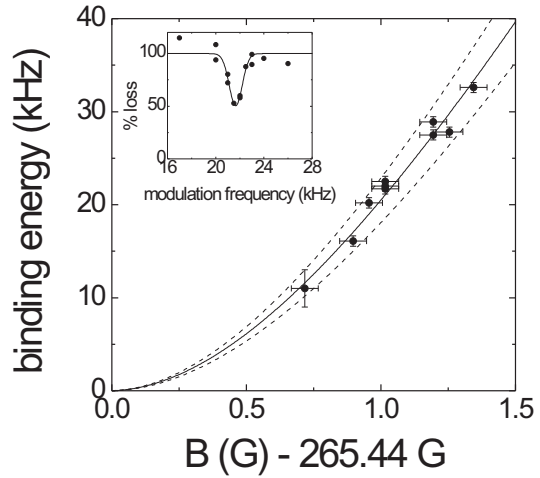}}
    \caption{Resonant frequency of atom loss as a function of magnetic field.
    The solid line is a fit to the data based on the universal
    binding energy of $s$-wave Feshbach molecules and the dashed lines
    represent the uncertainty in the Feshbach resonance width. (Inset) An atom-loss
    spectrum at 266.5 G as a function of the modulation frequency. The
    modulation converts roughly 50$\%$ of the gas to molecules.
    The loss is centered at 21.7 kHz with a width of $0.6\pm0.2$ kHz.
    The solid line is a gaussian fit to the data.  We
    report the uncertainty in the binding energy as the width of
    the loss spectrum because we lack a detailed understanding of the
    lineshape.
    \label{bind}}
    \end{center}
    \end{figure}

We created heteronuclear molecules using a small oscillating magnetic field to do spectroscopy on
the Feshbach bound state. The oscillating field causes two atoms to bind together and form a
molecule \cite{Thompson2005b,Bertelsen2006a}. We first ramp the magnetic field from 269 G to a
selected value between 266.2 and 267 G in 0.3 ms. The field modulation is then applied using the
Helmholtz coil pair for 20 ms with a peak-to-peak amplitude between 0.6 and 1.0 G at a frequency up
to 40 kHz. Next the field is returned to 269 G and held there for 5 ms, allowing any molecules made
during the modulation to decay. Finally we determine the number of atoms remaining as before. As in
Ref. \cite{Thompson2005b}, we observe strongly enhanced atom loss at certain frequencies, and the
loss depends on the duration and amplitude of the modulation. A typical loss spectrum is shown in
the inset of Fig. \ref{bind}; the frequency at which we observe maximum atom loss gives a measure
of the binding energy of the molecules \cite{Papp2006a}.

The resonant frequency of maximum atom loss is shown in Fig. \ref{bind} for various magnetic fields
near the Feshbach resonance. For each loss curve the amplitude used converted roughly half the \ra
gas into molecules. The solid line is a fit to the data based on the universal form of the
molecular binding energy near an $s$-wave Feshbach resonance \cite{Kohler2006a}. In the fit the
background scattering length is fixed to the value $213\pm7$ $a_0$ \cite{Burke1999a}, and the
Feshbach-resonance peak position and width are varied; the best fit finds the peak position and
width to be $265.42\pm0.08$ G and $5.8\pm0.4$ G, respectively. These results are consistent with
our previously discussed determination of the Feshbach-resonance peak position and with the
predicted width of the Feshbach resonance in Refs. \cite{Burke1998a,Bohn2006a,Kokkelmanns2006a}.

In summary, we have created heteronuclear Feshbach molecules from an ultracold gas of \ra and
$^{87}$Rb.  We demonstrated that molecules can be produced with two methods, magnetic-field sweeps
and resonant-field modulation. The conversion efficiency of \ra into molecules can reach $60\%$
even when that gas is not quantum degenerate.  The heteronuclear molecules described here are
ultracold and are stable for at least 1 ms.  These conditions may provide a first step toward the
efficient production of ground-state heteronuclear molecules.

Recently evidence of heteronuclear molecule creation has been reported using $^{40}$K and $^{87}$Rb
in Refs. \cite{Jin2006a,Ospelkaus2006a}.

We gratefully acknowledge useful discussions with Debbie Jin, John Bohn, Servaas Kokkelmans, and
Cindy Regal. We thank Josh Zirbel and Juan Pino for experimental assistance.  S. B. P acknowledges
support from an NSF Graduate Fellowship. This work has been supported by NSF and ONR.

\bibliographystyle{prsty}
\bibliography{sp_refs}

\begin{thebibliography}{10}
\bibitem{Papp2006a}
{T}he atom loss spectrum was recorded for a \rb gas with 50$\%$ condensate
  fraction and two non-condensed gases with temperatures of 130 nK and 190 nK.
  Over this range in temperature the measured resonant frequency shifts by only
  5(2) Hz/nK. The small temperature dependence that we observe can be explained
  by the fact that the molecule creation process preferentially selects atom
  pairs with low relative momentum and there exists an abundance of pairs
  available for each \ra atom in the gas.
\end{thebibliography}

\end{document}